\documentclass[twocolumn,aps,prl,amsmath,amssymb,showpacs,preprintnumbers]{revtex4}
\begin{document}

\title{Comment on ``Quantum bounce and cosmic recall''}

\author{Martin Bojowald}
\affiliation{Institute for Gravitation and the Cosmos,
The Pennsylvania State University, 104 Davey Lab, University Park,
PA 16802, USA}


\begin{abstract}
 A recently derived inequality on volume fluctuations
 of a bouncing cosmology, valid for states which are semiclassical
 long after the bounce, does not restrict pre-bounce fluctuations
 sufficiently strongly to conclude that the pre-bounce state was
 semiclassical except in a very weak sense.
\end{abstract}

\maketitle

The authors of \cite{BounceRecall} consider a loop quantum
cosmological model for a free, massless scalar field $\phi$, which is
known to bounce without a big bang singularity. The question of
interest is whether a generic state {\em assumed only to be
semiclassical at late times} must have been semiclassical at early
times.  An inequality is derived for the difference of relative volume
fluctuations {\em squared},
\begin{equation} \label{D}
 D:=\left|\lim_{\phi\to\infty} 
\left(\frac{(\Delta V)(\phi)^2}{\langle\hat{V}\rangle(\phi)^2}- 
\frac{(\Delta V)(-\phi)^2}{\langle\hat{V}\rangle(-\phi)^2}\right)\right| < 
8\frac{\Delta b}{\langle\hat{b}\rangle}
\end{equation}
where $\phi$ is a measure of time and $b$ a curvature parameter
conjugate to $V$. Eq.~(\ref{D}) shows that relative volume
fluctuations can only change by a very small value in absolute terms.
To specify this further, the authors consider a numerical example in
which relative curvature fluctuations are of the size $10^{-57}$ as
the amount by which squared volume fluctuations can change. From this
observation, the authors state ``(an almost) total recall.''

Implications of (\ref{D}) depend on the notion of semiclassicality
used, which is only a very weak one in \cite{BounceRecall}. To analyze
this properly, it is useful to ask what (\ref{D}) tells us about the
{\em relative} change in relative volume fluctuations. Absolutely,
$\Delta V/\langle\hat{V}\rangle$ cannot change much. But the size of
relative volume fluctuations is very small in the first place, and
adding a small number such as $10^{-57}$ to something possibly even
smaller could mean a significant change. Also the degree of smallness
matters, not just smallness compared to one. Here, it becomes
important that on the left hand side of (\ref{D}) we have the relative
fluctuations {\em squared}, but not on the right: For
$\Delta b/\langle\hat{b}\rangle\sim \Delta V/\langle\hat{V}\rangle$,
each term on the left of (\ref{D}) is smaller than the right hand
side, not just the difference.

In their example, the authors use the more specific assumption that,
in the late-time semiclassical state, $\Delta V/\langle\hat{V}\rangle$
is nearly equal to $\Delta b/\langle\hat{b}\rangle$, with minimal
uncertainty product ($\alpha_1\sim 1$). This gives rise to a crucial
inconsistency in \cite{BounceRecall} because the authors use here a
condition much stronger than their understanding of semiclassicality
elsewhere.  At late times $\sqrt{G}\phi\gg1$ we have $\Delta
V/\langle\hat{V}\rangle\sim 10^{-57}$. The first term in $D$ is then
$10^{-114}$, but $D$ is bounded by the much larger $10^{-57}$. If the
inequality can be saturated, $\Delta V/\langle\hat{V}\rangle$ before
the bounce, at $-\sqrt{G}\phi\gg 1$, could be as large as
$\sqrt{10^{-57}}\approx 10^{-28}$ which is huge compared to the value
at late times.  The uncertainty product is then far from minimal
($\alpha_1\sim 10^{28}\gg 1$) unlike at late times, which invalidates
the conclusion of \cite{BounceRecall}: Reproducing a number up to a
factor of $10^{\pm 28}$ hardly constitutes ``(an almost) total
recall.''

After the bounce only semiclassicality is supposed to be assumed, but
a minimum uncertainty product is actually used.  There are only two
consistent options to conclude: (i) the early state was {\em not
semiclassical} because its uncertainty product can be much larger than
minimal, or (ii) it is considered {\em semiclassical by weaker
standards}. In the second case one has to justify the much stronger
late-time assumption $\alpha_1\sim 1$ --- or relax it.  A
justification has not been provided; and if it is relaxed to allow
deviations from minimal uncertainty by a factor up to $\alpha_1\sim
10^{28}$, the early uncertainty product would be allowed to be even
larger (by $10^{42}$).  Again, this is much weaker than the new
late-time condition. Comparable pre- and post-bounce uncertainty
products are implied by (\ref{D}) only if $\Delta
V/\langle\hat{V}\rangle\sim 1$, which is {\em not semiclassical}. The
analysis in \cite{BounceRecall} regarding semiclassicality is
intrinsically inconsistent due to different standards used, which
shows that the role of semiclassicality of states is much more subtle
than it may appear.

In \cite{Harmonic}, by contrast, dynamical coherent states provide
bounds for the {\em relative} change of relative volume fluctuations,
exploiting the availability of a solvable model
\cite{BouncePert,BounceCohStates}. Coherence here means that the
uncertainty relation is aleays exactly saturated.  Thus, the
assumption is stronger than semiclassicality, allowing more control.
Still, relative changes even in such a state are bounded only by a
factor of around $20$. This is much smaller than the numbers
controlled by (\ref{D}), but still more than one as one should have it
for what one could call a recall.  This is the basis of cosmic
forgetfulness \cite{BeforeBB}: not all the fluctuations (and higher
moments) of a state before the bounce can be recovered after the
bounce, and values depend very sensitively on the late-time state.

\end{document}